\documentstyle[aps,prb,multicol,epsfig]{revtex}

\begin{document}
\title{Effects of   magnetic field induced chiral-spin interactions on  quasi-one-dimensional spin  systems
}
\author{R. Chitra$^{1}$ and R. Citro$^{2}$}
\address{$^{1}$Institut Laue-Langevin, BP 156 Grenoble 38042, France\\
$^{2}$Dipartimento di Scienze Fisiche ''E.R. Caianiello''\\
Universit{\`a} di Salerno and Unit{\`a} INFM di Salerno, Italy}
\date{\today}
\maketitle

\begin{abstract}
It is known that in  certain non-bipartite
 quasi-one dimensional spin  systems in a magnetic field,
in addition to the usual Pauli
coupling of the spins to the field,
new parity breaking three spin interactions, i.e. chiral
spin interactions, are induced  at higher order
due to virtual processes  involving
 the  intrinsic electronic nature of the underlying spins.
The strength of these
interactions depend strongly on the orientation of the field, a feature
which can be exploited to detect chiral effects
 experimentally. In many spin systems, these chiral interactions are generated
and
should be taken into account before any comparison with experiments
can be made.     We study the effect
of the chiral interactions on certain quasi-one dimensional gapped spin
half systems and show that they
can potentially alter the physics expected from the Pauli coupling alone.
In particular, we demonstrate that these terms alter
the universality class of the
C-IC transition in spin-tubes. More interestingly,
in weakly coupled XX zig-zag ladders,
 we find that the field induced chiral term
can close the singlet gap and drive a second order transition in the
non-magnetic singlet
sector, which then manifests itself as a two component Luttinger
liquid-like behaviour in the
spin correlation functions. Finally, we discuss the
relevance of our
results to experiments.
\end{abstract}

\bigskip

\section{introduction}

\label{intro}

Quasi-one dimensional quantum spin systems have long been known to exhibit a
spectrum of behaviours, ranging from the   gapless critical behaviour in spin-$\frac1{2}$ integer
systems, to  gapped spin Peierls systems and  Haldane phases in integer spin
 systems.
The effect of an external magnetic field is usually described
through the Pauli
coupling (${\bf h} \cdot {\bf S}$) of the spins to the field.
These fields can have dramatic effects on the spin systems. For example,
in gapped spin systems, the  magnetic field induces
 the  commensurate-incommensurate  (C-IC) transition, where the
 IC phase is characterized by  gapless spin
excitations described by a Luttinger
liquid and the   corresponding spin
 correlation functions exhibit power law behaviours whose exponents,
however, vary from
system to system.
 However, it is important to remember that in real materials, the spins are
electrons whose charge degrees of freedom are frozen by
strong electron-electron
interactions.
 In a physical system, the magnetic field couples not only to the spin degrees of freedom, via the Pauli term, but also to the orbital degrees of
freedom of the electrons.  This results in non-trivial additions to the
effective spin hamiltonian.
In fact, it is known
 that in the limit of large Coulomb
repulsion, this coupling can in principle  induce a three spin chiral
interaction between neighbouring spins \cite{chitra95,rokhsar}.
Such a term arises in two dimensional systems on
the triangular and Kagome  lattices.  These chiral terms also arise in certain  quasi one
 dimensional systems like zig-zag ladders, frustrated
ladders, spin tubes, saw tooth chains, etc, whose geometries permit the formation of a closed plaquette involving three neighbouring
spins.
The possible generation of such chiral terms in  quasi-one dimensional spin systems stemming from
the underlying electronic nature of realistic models has been
systematically overlooked in the literature.
In this paper, we show that these chiral terms are indeed relevant and can
change the  behaviour of certain spin   systems like spin-tubes and zig-zag ladders in   magnetic
fields.
The possible effects engendered by these terms should be
studied before  sensible comparisons with
 experiments in a magnetic field can be made.
 Another important motivation is
that the effects of these chiral interactions depend strongly
on the direction of the applied field, and could be detected
experimentally in measurements of various quantities ranging
from specific heat, NMR rates and dynamical structure factor measurements.

One class of systems, where the chiral term will surely be relevant
are systems which support
chiral excitations in zero field.
Examples are the recently studied
 zig-zag spin $s$ ladders \cite{nersfru,azaria,andrei}  where certain parity breaking interactions
induce spin nematic ground states in zero magnetic field and
spin tubes \cite{citro_SU(3)},  whose ground states  have
chiral properties in the
absence of a magnetic field.
These spin tubes  were also shown to  exhibit an interesting C-IC
transition where the IC phase is described
by a two
 component Luttinger liquid
 describing gapless spin and {\it chiral} excitations.
 In these systems,  the new  field induced chiral term can completely
alter the physics expected from the Pauli coupling alone.
 Normally, since we do not know of physical probes which couple
directly to the  chirality, it is
quite hard to  perceive  experimentally the  chiral aspects of the ground state.
 However, the competition between the chiral term  and the Pauli coupling of the  spins  to an external magnetic field
 can be exploited to obtain experimental signals of chirality in certain systems.
 The main
objective of this paper is to study the influence of these
chiral terms on the gapless incommensurate phases of certain spin
systems.
 These are especially
pertinent given the surfeit of experiments on spin systems in strong
magnetic fields\cite{chaboussant,spintube}.

The paper is organised as follows: we first outline the derivation
of the chiral spin term starting from the Hubbard model\cite{chitra95}.
We then study the
effect of this term on the spin-tube
and other quasi-one dimensional systems, in the limit of strong planar
interchain couplings. We then  discuss  the very interesting
example of the effects of the chiral interaction on the weakly coupled
 zig-zag ladder, where the magnetic field induced chiral interaction
 triggers a transition in the spin singlet sector.
 Finally, we conclude with a discussion of the
experimental consequences of this interaction.

\section{chiral interaction}

\label{}

In the presence of an external magnetic field, it is customary to only
consider the Pauli interaction. However, this completely ignores the
coupling of the field to the orbital part of the motion  which can result
in non-trivial corrections to the effective spin hamiltonian.
We start by considering a system of interacting electrons
described by a Hubbard
model
\begin{equation}
H=-\sum_{\langle ij \rangle } t_{ij} [c^{\dagger}_{i}c_{j}+ h.c.] +
U\sum_{i}n_{i\uparrow} n_{i\downarrow},  \label{elham}
\end{equation}
\noindent
where the first term describes the  hopping of electrons with
spin $\sigma$ from sites $i$ to $j$  and
the second, the on-site Coulomb repulsion
between electrons. The Pauli coupling of the electronic spins ${\bf S}$
to the external
 magnetic field ${\bf h}$  is given by $\mu_0g{\bf h}\cdot{\bf S}$, where $\mu_0=|e|\hbar/2mc$
 is the Bohr magneton and the gyromagnetic ratio $g\simeq 2$. Hereafter we set
 $g\mu_0=1$.
The magnetic field also couples to the electronic motion via the hopping matrix elements
$t_{ij}$, which   become complex, i.e., it
picks up a phase $t_{ij}\to t_{ij}\exp [ie \int_i^j {\bf A}\cdot d{\bf r}]$,
where, ${\bf A}$ is the vector potential associated with the external field $%
{\bf h}$. If the electron hops around a closed loop, we immediately see that
the net hopping matrix element is proportional to the magnetic flux enclosed by
the loop. Since an external magnetic field explicitly breaks time reversal
invariance, it is interesting to study whether time reversal violating terms
are generated in the spin hamiltonian in the large $U$ limit. At second order  in perturbation
 theory,  no   parity breaking
 terms are generated and one obtains  only the Heisenberg term ${\bf S}_i\cdot {\bf S}_j$.
Chiral terms, can therefore, be obtained only in higher order perturbation theory.  An example of
 a chiral term which respects all spin symmetries and yet  violates time reversal and parity,  is the triple product of three spins ${\bf S}_i.
{\bf S}_j\times {\bf S}_k$.
 Is such a term generated within the
Hubbard model with nearest neighbour hopping $t$ in the presence of an
external magnetic field ? At half filling, it can be shown that no parity and time-reversal
violating terms are induced on bipartite lattices \cite{chitra95}. However, on non-bipartite
lattices, such a term is explicitly generated at order $U^{-2}$ in
perturbation.
\begin{equation} \label{cchi}
H_{chiral}= {\frac{{24 t^3} }{{U^2}}} \sin(e \Phi)
 \sum_{\langle ijk\rangle} {\bf S}_i. {\bf S}_j\times {\bf S}%
_k
\end{equation}
where the summation is over sets of nearest neighbour sites which form
triangles. We have restricted ourselves to nearest neighbour hopping $t$ in the above.
$\Phi$ is the flux enclosed by the plaquette formed by the sites $i,j$ and $%
k $. In terms of the   magnitude of the field $h$,   $\Phi= Ah \cos (\theta)$ where $A$ is the area of the
plaquette and $\theta$ is the angle made by the magnetic field with respect
to the normal to the plane of the plaquette.  Substituting this in  (\ref{cchi}), we see  that
no chiral term is generated for magnetic fields applied
parallel to the plane of the plaquette. Additionally,   the  strength of the chiral term
  varies from a minimum to a
maximum as a function of the direction of the applied field. For
real materials, the typical area enclosed by these triangles and hence the
enclosed flux being very small, the $\sin(e\Phi)$ term in (\ref{cchi}) can be
expanded to yield
\begin{equation}  \label{chiham}
H_{chiral}= {\frac{{24 t^3} }{{U^2}}} eA h \cos{\theta} \sum_{\langle
ijk\rangle} {\bf S}_i. {\bf S}_j\times {\bf S}_k
\end{equation}

Since the chiral term is  generated at higher order, its strength   is lesser than
the Heisenberg exchange as well as the Pauli interaction. Despite this, it was shown
\cite{chitra95} that in certain cases the
chiral term may have a larger effect  on the ground state than the Pauli coupling of electron
spins to the external magnetic field, thereby  drastically  affecting the ground state
properties of the model.
One example is the
case where the ground state of the system in zero field is doubly
degenerate with opposite chiralities. Then, when an external
magnetic field is applied, in
addition to the effects of the Pauli term, the chiral degeneracy is broken
by (\ref{chiham}). The effect could be more pronounced in the case where the
ground states are degenerate chiral {\it singlets} because  here, though the chiral term
still breaks the degeneracy,  the Pauli interaction term has no effect on
the ground state for  fields smaller than the singlet-triplet gap.

Since, no  chiral interactions
are generated for fields  applied  parallel to  the plane of the triangular
plaquettes, the behaviour of the system
could, in principle,  be  different depending on whether the field is applied
parallel or perpendicular to the plane of the system.  This is especially relevant for the critical
behaviours in the IC phase, where exponents could show a variation with the
field direction.
Such a behaviour, provides us with
a physical handle for determining whether an isotropic spin system is
quasi-one dimensional or not i.e., if the  behaviour  is
different for the two directions of the field, one can presume that this
difference stems from chiral interactions which necessarily require the
system to be atleast quasi-one dimensional!

\section{ Strongly Coupled spin tubes and ladders}

\label{}

In this section, we discuss the effects of the chiral term
 on the IC phase of certain  strongly coupled spin-gap systems, whose geometry allows for
the existence of chiral terms in a magnetic field.   In the absence of chiral interactions,
 the behaviour of these
systems in the IC phase,  has been well studied and
catalogued in the
literature\cite{nersfru,citro_SU(3),mila_ladders,white,mikeska_dim,sen}.
As we will show below, the chiral term does not necessarily alter the nature
of the incommensurate phase always. The effects of the chiral
interaction are drastic or trivial depending on the nature of the ground and excited
states of the system.

\subsection{Spin tubes}

\begin{figure}
\centerline{\epsfig{file=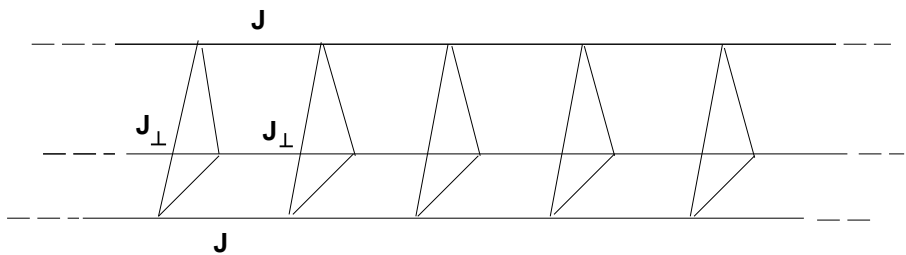,angle=0,width=12cm}}
\vspace{0.5 cm}
\caption{Cylindrical three-leg ladder (spin-tube).}
\label{fig:tube}
\end{figure}

The first  example we consider is  the strongly coupled
spin tube (i.e. a three-leg ladder with periodic boundary conditions)
studied in Ref.\onlinecite{citro_SU(3)}. As already mentioned
spin tubes, exhibit
a new class of C-IC behaviour. Starting from the
limit of strong-rung coupling (see Fig.\ref{fig:tube}),
the low-energy effective Hamiltonian (LEH) describing
the
gapless incommensurate  phase with finite magnetization
was derived in Ref.\onlinecite{citro_SU(3)}. This low energy Hamiltonian was
that
of a {two-component Luttinger liquid } describing {\it gapless spin and chiral
excitations} in the system.
The existence of a two-component Luttinger liquid phase
has important physical consequences. Namely, it implies that
both chiral and magnetic excitations contribute to a
non-zero magnetic susceptibility and
a strongly enhanced $T$-linear specific heat.
This different behaviour stems basically from the presence of
a degenerate ground state with different chiralities, within an isolated
triangle.
 This
immediately poses the interesting question, what is the effect of the chiral interaction (\ref{chiham}) on the two component Luttinger liquid behaviour?
Clearly, we
expect  the chiral term (\ref{chiham}) to induce dramatic changes,
because it will break the
degeneracy of the ground state.
 To see this, we derive
the low-energy effective Hamiltonian for the spin tube in
the presence of the chiral interaction (\ref{chiham}). We will use this
effective theory to discuss the physics of the IC phase and
the  resulting spin-spin correlation functions.

The Hamiltonian of the  spin tube
 in the presence of an external magnetic field ${\bf h}$ and a spin-chiral
interaction is given by,
\begin{equation}
H=\sum_{i=1}^{N}\sum_{p=1}^{3}({J{\bf{S}}_{i,p}\cdot{\bf {S}}_{i+1,p}+J_{\perp }\cdot%
{\bf{S}}_{i,p}\cdot{\bf{S}}_{i,p+1}}{\rm {\ }{-{\bf{h}}\cdot{\bf{S}}_{i,p})+\mu_c h
\sum_{i=1}^{N}{\bf{S}}_{i,1}\cdot({\bf{S}}_{i,2}\times {\bf{S}}%
_{i,3})},}  \label{ham}
\end{equation}
\noindent where  $(i,p)$ denote the  site and chain indices  respectively \
 and periodic
boundary conditions imply that the site $%
(i,4)$ is identified with the site $(i,1)$.  The first two terms represent the spin Hamiltonian for
the spin tube, the third the Pauli coupling of the spins to the magnetic
field and the last
term is the chiral term derived earlier. $J$ is the
coupling along the chain, $J_{\perp }$ the transverse coupling between the chains
and, $\mu_c=
 {\frac{{24 t^3} }{{U^2}}} eA  \cos{\theta}$
 is the strength of the
spin-chiral interaction.
As in Ref.\onlinecite{citro_SU(3)}, we consider the limit $J_{\perp }\gg J$. In this limit, the properties of
the system can be studied by perturbing in $J$ around the  limit of decoupled triangles.
For $J=0$, the system consists of
independent triangles. The eight states of a given triangle split  into  two spin-$\frac1{2}$
doublet ground states, with energy $E_{s}=-3J_{\perp }/4$ and an excited
spin-$\frac3{2}$ quadruplet, with energy $E_{t}=3J_{\perp }/4$. The spin
doublet
ground state has chiral eigenvalues  $\pm$.
For  strong magnetic fields applied parallel to the plane of the triangle,
no chiral terms are generated. Consequently, the degeneracy in the ground
state is only partially lifted and one reverts back to the case
studied in Ref.\onlinecite{citro_SU(3)}. On the other hand, for fields
 perpendicular to the plane of the triangle,
the ensuing Pauli and  the spin-chiral terms lift all the degeneracies.
 For instance, for the spin doublet states,
 the Pauli and chiral terms
contribute $\pm h/2$ and $\pm \mu_c \sqrt{3}/4$ respectively,
to the energy . Note,
however, that the chiral
term has no effect on the ferromagnetic spin $\frac{3}{2}$ excited states.
This is expected since the chiral term, which is nothing but a  triple
product,  has a non-zero expectation value only for  non-collinear spin configurations.
Therefore, for  the field  applied in the $z$ direction, the resulting energies are
\begin{eqnarray*}
{\tilde{E}}_{t} &=&E_{t}-hm_{3/2} \\
{\tilde{E}}_{s} &=&E_{s}-hm_{1/2}\mp {\frac{{\sqrt{3}\mu_c h}}{4}}
\end{eqnarray*}
Here $m_{s}$ just refers to the $S_{z}$ value of the state considered.
There exists a
critical field $h_c$, such that for $h<h_c$ the ground state has magnetisation
 $m=\frac1{2}$ and for
$h>h_c$ the ground state has $m=\frac3{2}$. In
the absence of chiral interactions, the critical field $h_{c}$ required to
close the gap is given by $h_{c}=E_{t}-E_{s}=3J_{\perp }/2$. However, in the
presence of chiral interactions, the critical field is slightly increased
and is given
\begin{equation}
h_{c}={\frac{{(E_{t}-E_{s})}}{{1-{\frac{{\sqrt{3}\mu_c }}{4}}}}}\equiv {\frac{{%
3J_{\perp }}}{{2-{\frac{{\sqrt{3}\mu_c }}{2}}}}}
\end{equation}
We therefore see that the chiral interaction
changes the magnitude of $h_{c}$
in a subtle way. Note also that the magnitude of the
critical field depends on the direction of the applied field. It is minimum
for a field applied in the plane of the triangle i.e., $x$ direction where $%
h_{c}=3J_{\perp }/2$ and  is maximum for  the field in
the $z$ direction.
A small non-zero value of the coupling $J$, broadens the abrupt transition
from  a state with $m=\frac1{2}$ to $m=\frac3{2}$. This results in two critical fields $H_{c1}$ (where the
$m=\frac{1}{2}$ state stops being the ground state ) and $H_{c2}$ (where the
magnetisation saturates at $m=\frac{3}{2}$ corresponding to a ferromagnetic
state)  with the magnetisation varying smoothly in between.
The  difference $H_{c2}- H_{c1} $ is of the order of the
coupling $J$. We now write an effective hamiltonian describing the low
energy physics of the spin tube close to the critical magnetic field.
Since at $h_{c}$ the $m_s=\frac{3}{2}$ state becomes
degenerate with only one of the chiral states with $m_s=\frac{1}{2}$ we retain
only these two states per site :
\begin{eqnarray}
\alpha _{\frac{3}{2}} &=&|\uparrow \uparrow \uparrow \rangle \\
\alpha _{\frac{1}{2}} &=&{\frac{1}{\sqrt{3}}}[|\downarrow \uparrow \uparrow
\rangle +j|\uparrow \downarrow \uparrow \rangle +j^{2}|\uparrow \uparrow
\downarrow \rangle ]
\end{eqnarray}
where $j=\exp (2\pi i /3)$. We now introduce a pseudo-spin half operator ${\bf %
T}$ which acts on the subspace containing the above two states. By
inspection, we can rewrite the original spin operators ${\bf S}$ in terms of
${\bf T}$ i.e.,
\begin{eqnarray}
S_{i,p}^{+} &=&{\frac{{j^{p-1}}}{\sqrt{3}}}T_{i}^{+} \\
S_{i,p}^{z} &=&\frac{1}{3}(1+T_{i}^{z})  \label{ops}
\end{eqnarray}
\noindent Using this, the effective hamiltonian, to first order in the
coupling between the triangles is now given by
\begin{equation}
H_{{eff.}}=\frac{J}{2}%
\sum_{i=1}^{N}(T_{i}^{+}T_{i+1}^{-}+T_{i}^{-}T_{i+1}^{+})+\frac{J}{3}%
\sum_{i=1}^{N}T_{i}^{z}T_{i+1}^{z}-h_{{eff.}}\sum_{i=1}^{N}T_{i}^{z}
\label{heff}
\end{equation}
\noindent where $h_{eff}=h-h_{c}-{\frac{{2J}}{3}}$ and contains the
effects of chirality. The previous derivation is equivalent
to that of the low-energy effective Hamiltonian for three coplanar
spin-$\frac1{2}$
chains considered in Ref.\onlinecite{sen}. The model (\ref{heff}) is
 the simple XXZ chain in a magnetic field. Since the coefficient
of the $z$ component of the interaction is smaller than that of the
transverse components, we see that the above system corresponds to that of
the XXZ chain in its gapless phase. A lot is known about the model and
results for the exponents of various correlation functions as a function of
the applied field can be obtained from Ref.\onlinecite{haldane}.

Using  the Jordan-Wigner transformation  we  fermionize  (\ref
{heff}), to obtain
\begin{equation}
H_{{\rm {eff.}}}=\frac{J}{2}\sum_{i}[c_{i}^{+}c_{i+1}+h.c.]+\frac{J}{3}%
\sum_{i}n_{i}n_{i+1}-\mu \sum_{i}n_{i}+{\rm const.}  \label{heffferm}
\end{equation}
$\mu =h-h_{c}-J/3$. The values of the critical fields $H_{c1}$ and $H_{c2}$
for the spin-tube can now be obtained from the effective model (\ref
{heffferm}). When $h=H_{c1}$, the band of spinless fermions is empty. This
occurs when the chemical potential $\mu $ equals the band minimum i.e.,
when $\mu =-J$ which in turn yields $H_{c1}=h_{c}-2J/3$. Similarly, $H_{c2}$
corresponds to a completely filled band of electrons or equivalently
an empty band of
holes. The simplest way to describe this situation is to perform a
particle-hole transformation on the Hamiltonian (\ref{heffferm}): $%
c_{i}^{\dagger }$ $\rightarrow h_{i}.$ Up to a constant, the new Hamiltonian
reads:

\begin{equation}
H_{{\rm {eff.}}}^{h}=-\frac{J}{2}\sum_{i}[h_{i}^{+}h_{i+1}+h.c.]+\frac{J}{3}%
\sum_{i}n_{i}^{h}n_{i+1}^{h}-\mu _{h}\sum_{i}n_{i}^{h}
\end{equation}
where the chemical potential $\mu _{h}=-\mu +2J/3$ . In terms of holes, $%
H_{c2}$ corresponds again to the chemical potential \ where the band stars
to fill up, thus we find $H_{c2}=h_{c}+2J/3$.

The Hamiltonian (\ref{heff}) can be easily bosonized to obtain the various correlation
functions and exponents in the incommensurate phase. Using
the standard expressions for bosonization
\begin{eqnarray}
T_{z}(x) &=&-\frac{1}{\pi }\partial _{x}\phi +(-1)^{\frac{x}{a}}\frac{\cos {%
2\phi (x)}}{\pi a} \\
T_{+}(x) &=&\frac{1}{\sqrt{2\pi a}}e^{-i\theta }[(-1)^{\frac{x}{a}}+\cos {%
2\phi (x)}]  \label{spins}
\end{eqnarray}
\noindent we obtain the following continuum theory
\begin{equation}
{\cal H}={\frac{1}{{2\pi }}}\int dx[uK(\pi \Pi )^{2}+({\frac{u}{K}}%
)(\partial _{x}\phi )^{2}]+{\frac{1}{{\pi }}}\int dxh_{eff}\partial
_{x}\phi ,  \label{bosham}
\end{equation}
where $\Pi $ is the momentum conjugate to the field $\phi $ where $u=\frac{%
\pi }{2}\frac{\sqrt{J^{2}-J_{z}^{2}}}{\cos ^{-1}(\frac{J_{z}}{J})}$ is the
spin wave velocity, and
\begin{equation}
2K=[1-\frac1{\pi}\cos ^{-1}(\frac{J_{z}}{J})]^{-1}
\end{equation}
This indicates clearly that in the presence of chiral interactions, the spin tube in the
IC phase,
 behaves as a {\it one-component Luttinger liquid},  describing spinless excitations alone.
 In the absence of the chiral term, both spin and chiral excitations were gapless which then led to
the two component Luttinger liquid  behaviour. Here, we see that the effect of the chiral term
 is to open a gap for the chiral excitations, which then  do not contribute
to the low energy behaviour.
The magnetic field
term, described by the gradient  in (\ref{bosham}), can be eliminated by a simple shift of the $%
\phi $ field i.e., $\phi \rightarrow \phi +\pi mx$ where the magnetisation
$m\propto h_{eff} K/u$.
For small values of
the field $h_{eff}$, the effective magnetisation ${m}$ increases
linearly with $h$. This field
shift  results in the appearance of incommensurate modes in the
spin-spin correlation functions.
Note that  $K$ varies from $K=1$ for the case of the XX- antiferromagnet
to  $K={\frac{1}{2}}$ for the isotropic XXX system.
 For the
present problem, $J_{z}=J/3$ resulting in $K=0.83$ and $u=1.2J$. It is well
known that for this value of $K$ the system is gapless and lies in the XXZ
universality class.
  However,  due to presence of marginal operators in the theory, the Luttinger parameter $K$  now
 varies
with the magnetisation. This variation depends also on the anisotropy $%
J_{z}/J=1/3$ and can be obtained  numerically using the Bethe ansatz
equations of  Ref.\onlinecite{haldaneexp}.

In the
critical incommensurate (IC) region characterized by a finite magnetization,
we can easily calculate the various correlation functions of the
effective spin chain and hence that of the original spin tube
\begin{eqnarray}
\chi _{p}^{zz} &=&\left\langle S_{p}^{z}(x,t)S_{p}^{z}(0,0)\right\rangle
\label{corrf1} \\
\chi _{p}^{\pm} &=&\left\langle
S_{p}^{+}(x,t)S_{p}^{-}(0,0)\right\rangle ,  \label{corrf2}
\end{eqnarray}
\noindent where $p$ is the chain index and $\pm$ refers to the intra and
interchain correlation, respectively. From (\ref{ops}), we
see that the magnetisation $\tilde m$ in the spin tube is related to
the magnetisation of the
 effective system by
 $\tilde m= (1-m)/3$.
 The first correlation  function is useful for
neutron scattering experiments, whereas the correlation function (\ref
{corrf2}) defines the staggered susceptibility that is useful for the
calculation of NMR relaxation rates.  Using  (\ref{ops}) which
 relate the original spins to the pseudo-spin $\frac{1}{2}$ variables, we obtain
\begin{eqnarray}
\chi ^{zz} &=&\frac1{9}\left[\left\langle T^{z}(x,t)T^{z}(0,0)\right\rangle +(1+ \left\langle
T^{z}\right\rangle )^{2}\right]  \label{chizz} \\
\chi _{p}^{+-} &=&\frac{1}{3}\left\langle
T^{+}(x,t)T^{-}(0,0)\right\rangle  \label{chistag}
\end{eqnarray}
\noindent
which in turn leads to the following results \cite{affleck_leshouches,chitra_externalh}:
\begin{eqnarray}
\chi_{p} ^{zz} &\simeq&\frac1{9}(1- m )^{2}+ \frac 1{9}\cos \pi
x(1-2m)(x^{2}-u^2t^{2})^{-K}+{ const} \ \  {K}\left( \frac{1}{(x-u t)^{2}}+%
\frac{1}{(x+u t)^{2}}\right) ,  \label{chiz} \\
\chi _{p}^{\pm} &\simeq&\frac1{3}\left[
(-)^{x/a}(x^{2}-u^2 t^{2})^{-\frac{1}{4K}}+\ const.\cos (2\pi
mx)(x^{2}-u^2t^{2})^{-(\frac{1}{4K}+K-1)}\right. .  \nonumber \\
&&\left. \left( \frac{e^{2i\pi m x}}{(x-u t)^{2}}+\frac{e^{-2i\pi m x}}{(x+u t)^{2}}%
\right) \right] {\rm {\ ,}}  \label{chipm}
\end{eqnarray}
where,  $u$ is the spin
mode velocity.  The correlation functions so
found show a power-law decay in space and time.
 As anticipated,
we see that two incommensurate modes appear at the
wave vectors $Q=2\pi (1-m),$ for $\chi ^{zz}$ and $Q=2\pi m$ \ for
$\chi_{p}^{+-}$, resulting in a different behaviour in the plane
perpendicular to the field and along the field.
The exponents of these power-laws now vary with the magnetisation $m$. These can be
obtained by numerically solving the Bethe ansatz equations of
Ref.\onlinecite{haldaneexp}.
 As the magnetisation
$m\rightarrow 0.5$ or equivalently when the average number of fermions goes
to $1$, the fermion band is completely filled and we expect to recover the
physics or free fermions whose $K=1$.

The behaviour found for the spin-spin correlation functions in the present
case, is wholly different from that found in Ref.\onlinecite{citro_SU(3)}.
\begin{eqnarray}\label{eq:general_correlation}
& &\chi _{p}^{\pm} \simeq \left(\frac
{1}{x^2 -(u_a t )^2}\right)^{\frac{K_a}{2}+ \frac1{2K_a}}  \cos (Q_1(x,t))
+ \cos(Q_2(x,t)) \left(\frac
{1}{x^2 -(u_a t  )^2}\right)^{\frac1{2K_a}} \left(\frac
{1}{x^2 -(u_b t  )^2}\right)^{\frac{K_b}{6}}+ \nonumber \\
& &\cos(Q_3(x,t)) \left(\frac
{1}{x^2 -(u_a t  )^2}\right)^{\frac{K_a}{8} +\frac1{8K_a}} \left(\frac
{1}{x^2 -(u_b t  )^2}\right)^{\frac{3K_b}{8}+\frac3{8K_b}}+\cos(Q_4(x,t))
\left(\frac{1}{x^2 -(u_a t  )^2}\right)^{\frac{K_a}{8} +\frac1{8K_a}}
\left(\frac{1}{x^2 -(u_b t  )^2}\right)^{\frac{3}{K_b}+\frac{K_b}{24}}
\end{eqnarray}
Here, $u_a,u_b$ are the velocities of the spin and chiral excitations and $K_a,K_b$
are the Luttinger parameters associated with these excitations. $Q_1,Q_2,Q_3$ and $Q_4$
are functions which fix the incommensuration. One of the primary differences between
(\ref{eq:general_correlation}) and (\ref{chipm}) is that the former
shows an explicit dependence
on two different velocities $u_{a}$ and $u_{b}$ related to the magnetic and
chirality modes and exhibits incommensuration at many more wavevectors. These velocities were
different and in addition to the sum of the power law correlations exhibited
by the two components, there is an interference between the two which
contributes a third term with a different exponent in the correlation
function for the spins. The two components will be distinguished when an
extra field  which couples to the spin current in the transverse direction is
applied.

To summarize, when the field is applied parallel to the plane of the triangles, no chiral interaction
is induced and the system exhibits the two component Luttinger behaviour
of (\ref{eq:general_correlation}). However, when   the field is applied along the length of the tube i.e.,
perpendicular to the plane of the triangles, the chiral interaction
plays a dominant role and we recover the
 single component behaviour
indicated in (\ref{chiz},\ref{chipm}).  The exponents of the various power laws in the correlation
functions will also be different in the
two cases. Moreover, the value of the critical fields $H_{c1}$ and
$H_{c2}$ are also different for the different field directions. These predictions are strong
signatures of the chiral term and they could
be verified experimentally.
Moreover, in the past there have been a lot of discussions of how to
experimentally establish the existence of chirality in systems. In spin
tubes and other systems, the change in critical behaviour can be perceived
as an indirect test of such chiral exchanges in realistic systems.

\subsubsection{Frustrated spin ladders}

\begin{figure}
\centerline{\epsfig{file=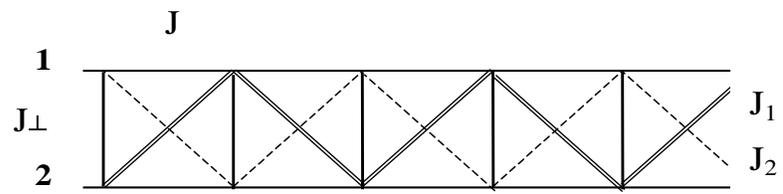,angle=0,width=7.2truein,height=9.2truein}}
\vspace{-7.5 cm}
\caption{Examples of frustrated ladders in the strong-rung coupling limit.}
\label{fig:frustl}
\vspace*{-9.5truecm}
\end{figure}

We have also analysed the effect of these chiral interactions on certain
strongly coupled frustrated ladders
considered by Mila\cite{mila_ladders} and shown in Fig.\ref{fig:frustl}.
In all of these cases, we find that the chiral term has no effect
whatsoever on the C-IC physics seen in these systems. This is due to
the fact that in each of the examples considered, the unperturbed ground state
(equivalent of the spin $\frac1{2}$ states on the triangles in the spin-tube)
has no interesting chiral properties i.e., they have no  low
lying chiral modes. As shown above, we introduce pseudo-spin operators
to derive the relevant effective hamiltonians close to $h_{c}$. In all these models,
it
turns out that the chiral term is effectively zero when written in
terms of these operators.
Consequently, the physics of these strongly coupled frustrated ladders  is completely unaffected by
the
chiral term.

\section{Weakly coupled zig-zag ladders}

In the previous section, we considered the effect of the chiral
interaction
on the spin-tube as an example of a strongly
coupled quasi-one dimensional systems. Here, the possible non-trivial effects
generated by the chiral term could be deduced by merely studying  the
the chiral properties of the relevant ``unperturbed''
ground state in the absence of the external field.
Such an approach fails for weakly coupled spin systems like
the zig-zag chains and frustrated ladders. There,
to begin with, one lacks a schematic picture of the ground state(s)
which is highly correlated. Neverthless, in such cases,  one can use
bosonization methods \cite{affleck_leshouches,gogolin}
to study the effects of the chiral terms on the long wavelength physics
of these systems.
Here, we  consider the example of the zig-zag ladder, where
the chiral terms are relatively simpler than those generated
in
frustrated ladders.
 We will restrict ourselves to
the limit of
very strong anisotropy that corresponds to two coupled XX chains. (We assume that the
anisotropies are generated by spin-orbit couplings, which do not affect the
form of the
chiral interaction.) The XX
zig-zag ladder was studied in Ref.\onlinecite{nersfru}  using a
mean-field treatment of the bosonized hamiltonian.
Here, we analyze the effect of
the chiral term on the mean-field solution via a similar treatment.

\begin{figure}
\centerline{\epsfig{file=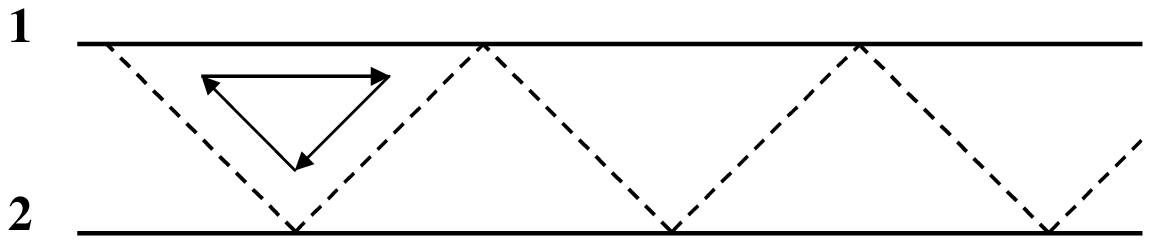,angle=0,width=6.2truein,height=9.2truein}}
\vspace{-7.5 cm}
\caption{Zig-zag ladder in the weak coupling limit. The arrows indicate the closed three-spins
path.}
\label{fig:zigzag}
\vspace*{-9.5truecm}
\end{figure}

 The zig-zag
ladder ( Fig.\ref{fig:zigzag})  has  spins ${\bf S}$ on chain $1$ and
spins ${\bf T}$ on chain $2$.
The
bosonized hamiltonian  for the  zig zag ladder has been derived by Nersesyan et al.\cite
{nersfru}.  The bosonic fields $\phi_{1}$ and $\phi_{2}$ and their
duals $\theta_{1}$ and $\theta_{2}$ are introduced
to
describe the  spins in the two chains, ${\bf  S}$ and ${\bf  T}$.
In terms of the the symmetric (antisymmetric) combinations, $\phi
_{s,a}=(\phi _{1}\pm \phi _{2})/\sqrt{2{\rm {\ }}}$ and  $\theta
_{s,a }=(\theta _{1}\pm \theta _{2})/\sqrt{2{\rm {\ }}}$ which
describe
the triplet (singlet) sector,
 the hamiltonian describing the long wavelength physics  of the anisotropic
XX zig-zag  ladder is given by
\begin{equation}
H=H_{0}+ H_{int}+ H_{mag}
\label{hamxx}
\end{equation}
where,
\begin{eqnarray}\label{hamterms}
H_{0}&=&\sum\limits_{\alpha =s,a }\frac{v}{2}\left[ (\partial
_{x}\theta _{\alpha })^{2}+(\partial _{x}\phi _{\alpha })^{2}\right] \nonumber \\
H_{int}&=&\gamma \partial _{x}\theta _{s}\sin \sqrt{2}\theta _{a}
\end{eqnarray}
and the Pauli coupling of the spins to an external field $h$
 (which affects only the
triplet sector) is given by
\begin{equation}
H_{mag}= h\frac{\sqrt[]{2}}{\pi}\partial_{x}\phi_{s}
\end{equation}
\noindent
 The spin-wave velocity $v\propto Ja$ and  $\gamma$
 is the
strength of the zig-zag coupling.   Note that the interaction term in
(\ref{hamterms})
is marginal.
Due to the anisotropy, we only consider the case where the field is
applied
along the
$z$-axis perpendicular to the plane of the zig-zag ladder. A magnetic field applied in the plane of the zig zag chain is very
complicated,
since even the simple Pauli coupling generates a gap in the triplet
sector.
In the absence of any field, the ground state of this model was shown \cite{nersfru}
to be a  spin nematic with gapless triplet
excitations and gapped singlet excitations. This state was also
characterized by    a power
law decay of the   staggered
transverse spin correlations at  an incommensurate wave vector $Q=\pi -\delta$, where $\delta$
depends on $v$ and $\gamma$.
A Pauli coupling to the magnetic field in the $z$ direction has no effect on
the staggered part of the correlation  and induces only a trivial incommensuration in the
uniform part of the spin correlation function. To the hamiltonian (\ref{hamxx})
we now add the chiral interaction term (\ref{hchiral})
involving the spins on every triangular plaquette. It takes the form
\begin{eqnarray}
&&  \label{hchiral} \\
H_{chiral} &=&\mu _{c}h({\bf S}_{i}-{\bf T}_{i+1})\cdot {\bf T}_{i}\times
{\bf S}_{i+1}
\end{eqnarray}
for every pair of spins ${\bf S_{i}},{\bf T_{i}}$.
In terms of the continuum fields,
\begin{eqnarray}
H_{chiral} &=&-\mu _{c}h\int dx\left[ \frac{1}{2\pi ^{2}a^{2}}\sin \sqrt{2}%
\theta _{a}(x)(1+2\sqrt{2}a\partial _{x}\phi _{a})+\right.
\label{hchiralbo} \\
&&\left. +\frac{1}{(2\pi )^{2}}(\partial _{x}\phi _{a}\partial _{x}\theta
_{s}-\partial _{x}\phi _{s}\partial _{x}\theta _{a})\right] ,
\end{eqnarray}
\noindent
We see that the first term in (\ref{hchiralbo}) is a relevant operator
of dimension one that can have the effect of eliminating the gap in the singlet sector. The other terms are all marginal.
The first two terms act exclusively on the singlet sector,  whereas the
last two terms couple the singlet and triplet sectors. Moreover, these
terms
will
compete with $H_{int}$ in (\ref{hamterms}).
 On
 physical grounds we speculate that the coupling of the chiral
interaction to the singlet sector arises from the hidden chirality
in the ground state.
The presence of so many
competing operators renders the problem highly complex. Here, since we
are
interested only in the qualitative effects of the chiral term,
we restrict ourselves to a simple self-consistent mean-field analysis which
leads us to the following Hamiltonian

\begin{equation}
H_{MF}=H_{s}+H_{a}
\end{equation}
\noindent with
\begin{equation}
H_{s}=\frac{v}{2}\left[ (\partial _{x}\theta _{s})^{2}+(\partial
_{x}\phi _{s})^{2}\right] +(h+\beta L_{3})\partial _{x}\phi _{s}+(\gamma
K_{1}-\beta K_{2})\partial _{x}\theta _{s},
\end{equation}

\begin{eqnarray}
\label{mfeq}
H_{a} &=&\frac{v_{}}{2}\left[ (\partial _{x}\theta _{a})^{2}+(\partial
_{x}\phi _{a})^{2}\right] +(\alpha _{1}+\alpha _{2}K_{2}+\gamma L_{2})\sin
\sqrt{2 }\theta _{a}+  \nonumber \\
&&+(\alpha _{2}K_{1}-\beta L_{2})\partial _{x}\phi _{a}+\beta L_{4}\partial
_{x}\theta _{a},
\end{eqnarray}
\noindent
where the mean-field parameters
\begin{eqnarray}
K_{1} &=&\langle \sin \sqrt{2}\theta _{a}\rangle {\rm {\ \ \
\ \ \ \ \ \ }}K_{2}=\left\langle \partial _{x}\phi _{a}\right\rangle
\label{one} \\
L_{1} &=&K_{2}=\left\langle \partial _{x}\phi _{a}\right\rangle {\rm {\ \ \
\ \ \ }}L_{2}=\left\langle \partial _{x}\theta _{s}\right\rangle   \label{two}
\\
L_{3} &=&\left\langle \partial _{x}\theta _{a}\right\rangle {\rm {\ \ \ \ \
\ \ \ \ \ \ \ \ \ \ }}L_{4}=\left\langle \partial _{x}\phi _{s}\right\rangle
{\rm ,}  \label{three}
\end{eqnarray}
and $\alpha _{1}=-\mu _{c}h/2\pi ^{2},$ $\alpha _{2}=-\sqrt{2}\mu _{c}h/\pi
^{2},\beta =\mu _{c}/2\pi ^{2}$. The  triplet channel ($s$) can be
solved easily by eliminating
the  terms linear in $\theta _{s}$ and $\phi _{s}$  through the
field shifts: $\theta _{s}\rightarrow \theta _{s}+(\gamma K_{1}-\beta
K_{2})x/2,$ and $\phi _{s}\rightarrow \phi _{s}+(h+\beta L_{3})x/2$.
 Analogously
in the $a$ channel, the gradient of $\phi_a$ can be eliminated by a simple shift
 $\phi _{a}\rightarrow \phi
_{a}+(\alpha _{2}K_{1}-\beta L_{2})x/2$.
 Self-consistency of the solution  then leads to the following relations
\begin{eqnarray}
 K_{2} &=&\frac{2\alpha_{2}+\beta \gamma }{\beta ^{2}-4}K_{1},{\rm {\ \
\ \ }}
L_{2} =\frac{1}{2} (\beta K_{2} - \gamma K_{1}) \\
L_3&=& -\frac1{2}(h+ \beta L_3)
\label{sfcon}
\end{eqnarray}

Here we have assumed that the ground state of the system is found
in the sector with nonzero  spin current $
\partial _{x}\theta
_{s }$.
>From the very structure of the mean field hamiltonian (\ref{mfeq}), we
 see that though triplet excitations remain gapless, the
situation
is not so clear for
 singlet excitations. The hamiltonian in the
singlet  sector takes the form
\begin{equation}\label{cic-ham}
H_{a} =\frac{v}{2}\left[ (\partial _{x}\theta _{a})^{2}+(\partial
_{x}\phi _{a})^{2}\right] +\mu \sin
\sqrt{2 }\theta _{a}+\beta L_{4}\partial
_{x}\theta _{a}
\end{equation}
\noindent
 where the coefficient of the sine term which generates a gap is given by
\begin{equation}
\mu =\alpha _{1}+\frac{2(\alpha _{2}^{2}+\alpha _{2}\beta \gamma +\gamma
^{2})}{\beta ^{2}-4}K_{1},  \label{mu}
\end{equation}

\noindent
Note that  the structure of $H_{a}$ is
similar to that of  systems with gapped triplet excitations
in an external magnetic field ( the
equivalent of the ``external'' magnetic field in this case being the
mean field parameter
$\beta L_{4}$).  The bare gap in this case is  $\Delta \propto \mu^{2/3}$.
>From the structure of the mean
field
equations,
we see that since $\mu$ depends on the external magnetic field, the gap
is a varying function of $h$.
 The
 field $L_{4}$ tends to reduce this gap.
Moreover, it is  known that these systems   exhibit the
C-IC transition when the strength of the field
attains a critical value which
equals the magnitude of the gap. For fields greater than the critical
field $h_c$ (not estimated here), the system has gapless excitations at incommensurate values of
the wave vector. We,  therefore, expect within mean field two kinds of solutions:
in the first one the gap induced by the sine term survives and, consequently, there
is no net "magnetisation" i.e., $\langle \partial_x \theta_a \rangle=0$. This implies
that $K_1 \neq 0$ and $L_3=0$.
 Using (\ref{sfcon}), we see that    $L_{4}= -h/2$ in the gapped phase.
The second solution describes the   "incommensurate" phase  with a finite "magnetisation"
$m_a$ i.e.,
$L_3=\langle \partial_x \theta_a \rangle =m_a$. In this phase, the effective field in the
singlet sector is given by $L_4 = -\frac{1}{2} (h+ \beta m_a)$.
One or the other    solution will be energetically favoured as the field is tuned.
In the gapless  IC phase,  the gradient in $\theta_a$ can be absorbed into the quadratic
part by the shift
 $\theta _{a}\rightarrow \theta
_{a}+\beta L_{4}x/2$;
 the singlet sector is  now effectively a Luttinger
liquid
described by
\begin{equation}
\label{lutts}
H_{a} =\frac{v_{a}(h)}{2}\left[ K_{a} (h) (\partial _{x}\phi_{a}
)^{2}+ \frac{1}{K_{a} (h)}(\partial
_{x}\theta_{a})^{2}\right]
\end{equation}
where the new velocity and the Luttinger liquid parameters $v_{a},K_{a}$
both
vary with the field. The results obtained earlier  in the context of the
normal C-IC transition\cite{chitra_externalh,heinz}
can be used to obtain $K_a(h)$. In fact,
$K_a(h=h_c)=2$.
In conjunction with the gapless spin excitations described by $H_s$,
this results in a two component Luttinger liquid like behaviour!
 The possibility of such a
transition,  however,
depends crucially on the mean field solutions.
The calculation of the regime of validity of the two solutions requires the knowledge of
the exact  expectation value of
the mass term
$\langle \sin \sqrt{2}\theta_{a} \rangle $
as a function of the prefactor of the  "magnetic field" like term
$\partial_{x}\theta_{a}$.
This expectation value is rather difficult to estimate and has only
been
calculated either for zero magnetic field or in the limit of  fields
larger
than the gap opened by the sine-Gordon term i.e., in the gapless phase where
$\Delta / \beta L_{4} \ll 1$
\cite{zamol_extfield,zamols}.
Here, we assume that the chiral term
has indeed driven the singlet sector across the transition into the
gapless ``incommensurate''phase and  solve for the mean field equations. In
this limit,
\begin{equation}
\left\langle \sin \sqrt{2} \theta _{a}\right\rangle =c\beta L_{4}\mu
 \label{mass}
\end{equation}
where  $c$ is some constant and $\mu $ given by (\ref{mu} ) is a function of the other mean
field
parameters.
Using (\ref{mass} ) and (\ref{three})  we can now solve explicitly
for $K_{1}$ to obtain
\begin{eqnarray}
K_{1} &=&\frac{\alpha _{1}}{f^{-1}(h)-\frac{2}{\beta ^{2}-4}(\alpha
_{2}^{2}+\alpha _{2}\beta \gamma +\gamma ^{2})}  \label{sce} \\
\mu  &=&\alpha _{1}+\frac{2\alpha _{1}(\alpha _{2}^{2}+\alpha _{2}\beta
\gamma +\gamma ^{2})}{f^{-1}(h)(\beta ^{2}-4)-2(\alpha _{2}^{2}+\alpha
_{2}\beta \gamma +\gamma ^{2})}.  \nonumber
\end{eqnarray}
where $f(h)= (ch)^{-1}$.
We find that the mean field solutions are self-consistent with the
limit
$\Delta/\beta L_{4} \ll 1$
used to obtain them. This limit  translates to the condition  $h \gg
\gamma/{\mu_{c}
}$ and  allows for a finite  $h_c$.

In conclusion, we have shown that, while the Pauli interaction with the
field
leaves the physics of the XX ladder completely unchanged,
at the mean field level the chiral term  radically alters the physics of
the
zig-zag chain.
  It generates a transition to a gapless regime in the singlet
sector for large enough values of the magnetic field.
The gapless singlet excitations manifest themselves in specific
heat
measurements and more importantly, in the spin correlation functions.
A new peak appears in the specific heat as a
function of the magnetic field.
Such a transition in the singlet space will result in a change of the exponents of
power laws in the transverse spin correlation functions, as we will see below.

\subsection{The correlation functions and relevance to experimental quantities}

We conclude the analysis of the effect of the chiral interactions on
the zig-zag chain with the calculation of the spin-spin correlation functions. They could be useful for
the neutron scattering experiments and the NMR relaxation rates measurements.
>From the previous mean-field solution we  get the following shifts of
\ the fields
\begin{eqnarray*}
\theta _{s} &\rightarrow &\theta _{s}+\frac{pK_{1}}{2v_{}}x \\
\theta _{a} &\rightarrow &\theta _{a}+\frac{p^{\prime }}{2v_{}}x \\
\phi _{a} &\rightarrow &\phi _{a}+\frac{\delta K_{1}}{2v_{}}x \\
\phi _{s} &\rightarrow &\phi _{s}+\frac{\delta ^{\prime }}{2v_{}}x
\end{eqnarray*}
\noindent
where
\begin{eqnarray*}
p &=&\frac{\gamma (\beta ^{2}-4)-\beta (\alpha _{2}+\beta /2\gamma )}{\beta
^{2}-4} \\
\delta &=&\alpha _{2}+\frac{\beta v}{2} \\
\end{eqnarray*}
for all values of the field $h$ and
\begin{eqnarray*}
p^{\prime } &=& 0 ;  {\ \ \ \ \ \ \ \ \ \ \ \ \ \ \ \ \ \ \ \ } \delta^\prime = h  { \ \ \ \ \ \ \ \ \ \ }       {\rm {for}}  { \ \ \ } h <h_c \\
p^{\prime } &=& -\frac1{2} (h+ \beta m_a);{\ \ \ } \delta^\prime = h+\beta m_a {\ \ \ }    {\rm for} { \ \ \ }  h>h_c
\end{eqnarray*}
\noindent
In  the C phase $(h <h_c)$, with gapless triplet excitations and
gapped singlet excitations,
 $\langle \theta_s \rangle =0$ and  $\theta _{a}$  has a non-zero expectation value  determined by the position of the
minima of the Sine-Gordon potential i.e., $\left\langle \theta _{a}\right\rangle =%
\sqrt{\pi /8}$sign$(\mu )$. This
allows us  to express the dual fields in the chains as
\begin{equation}
\label{theta-gap}
\theta _{1,2}=\frac{1}{\sqrt{2}}\theta _{s}^{0}(x)+\frac{pK_{1}
}{2\sqrt{2}v}x\pm \frac{\sqrt{\pi }}{4},
\end{equation}
Using (\ref{theta-gap}), the  staggered part of the transverse spin correlation functions are now
given by
\begin{equation} \label{transcor}
 \chi_{\pm}(x,t) \equiv \left\langle S_{1,2}^+  S_{1}^-\right\rangle \sim  (-1)^{x/a} e^{-iB
x}(x^2 - v^2 t^2) ^{-\frac{1}{8}}
\end{equation}
where $B=\left( \frac{pK_{1}}{2 {\sqrt 2}v}\right) $.
The transverse  correlations still fall off as a
power-law with exponent $1/4$, but they are incommensurate with  a  field
dependent characteristic vector:
\begin{equation} \label{incomc}
Q(h)=\pi -\left( \frac{pK_{1}}{2{\sqrt 2}v}\right) .
\end{equation}
Comparing (\ref{transcor},\ref{incomc})  with the results of Ref.\onlinecite{nersfru}, which corresponds to setting
$\mu_c=0$ in the above equations, we find that the exponents are
the same in both cases and the only effect of the chiral
term is to induce a magnetic field
dependent incommensuration. Moreover, the Pauli term
 does not
 induce any additional incommensuration in the staggered transverse spin correlation functions.
 Therefore, a varying $Q$ should serve as a good experimental probe of the  chiral
interactions.

In the gapless incommensurate phase of the singlet $(h>h_c)$, the singlet  excitations
 are described by a
Luttinger liquid (\ref{lutts}). In this case, both the gapless triplet and singlet excitations (which are probably
chiral) contribute to $\chi_{\pm}$.
Here, $\theta_{a}$  also has a zero expectation value  and the expression
(\ref{theta-gap})  is invalid in this phase. Instead,
\begin{equation}
 \label{theta-nogap}
\theta _{1,2}=\frac{1}{\sqrt{2}} (\theta _{s}(x)\pm
\theta_{a}^{} (x))
+\frac{pK_{1}\pm p^{\prime
}}{2\sqrt{2}v_{}}x.
\end{equation}
which leads to the following
asymptotic behaviour of the transverse
spin-spin correlation:

\begin{equation} \label{transcoric}
 \chi_{\pm}(x,t) \equiv \left\langle S_{1,2}^+  S_{1}^-\right\rangle \sim  (-1)^{x/a} e^{-iB_{\pm
}x}(x^2 - v^2 t^2) ^{-\frac{1}{8}} (x^2 - v_a^2(h) t^2) ^{-\frac{K_a(h)}{8}}
\end{equation}
where $B_{\pm }=\left( \frac{pK_{1}\pm p^{\prime }}{2{\sqrt 2} v}\right) $.
Note that $v$ is the velocity of the gapless triplet excitations and $v_a(h)$ is the
velocity of the gapless singlet excitations.
This  behaviour is indeed  reminiscent of a two component Luttinger liquid.
The equal-time transverse spin correlations  on the two chains fall off as a
power-law with exponent $K(h)= (1+ K_a(h))/4$, but they are incommensurate with  characteristic vectors:
\[
Q_{\pm}=\pi -\left( \frac{pK_{1}\pm p^{\prime }}{2{\sqrt 2}v}\right) .
\]
\noindent
 At the critical field $h=h_c$, since $K_a(h=h_c)=2$, the  spin correlation function
\begin{equation} \label{corrhc}
\chi_{\pm}(x,t)  \sim
 e^{i Q_{\pm}x}
(x^2 - v^2 t^2) ^{-\frac{1}{8}} (x^2 - v_a^2(h_c) t^2) ^{-\frac{1}{4}}
\end{equation}
  From (\ref{corrhc}) and (\ref{transcor}), we see that
the transition will manifest
itself
in the  equal time spin correlation function as an abrupt change in the exponent i.e.,
 {\it it jumps from}
$\frac1{4}$ to $\frac3{4}$  and then varies slowly with increasing field.
Such a dramatic change in the exponent could be easily seen
in  NMR and susceptibility measurements.

For strong zig-zag interactions $\gamma$, we anticipate that the chiral term  is irrelevant and  that
the    singlet sector remains gapped.
 It remains to be seen whether the transition in the
singlet space survives when fluctuations are taken into account. Also,
these are complicated sine-Gordon theories and one should worry about
solitons
and kinks.
An interesting question is whether such a physics survives in the
isotropic
case or when there is a finite $J_{z}$ coupling.

\section{conclusions}\label{}

Chiral interactions induced by the magnetic field appear in a variety of non-bipartite quasi one-dimensional spin systems. We have shown that these
terms could
dramatically alter  the physics of systems such as strongly
coupled spin tubes and weakly coupled zig-zag ladders. The effect of these chiral interactions depend strongly on the direction of the applied field and could be measured experimentally.
In the case of spin-tubes in a magnetic field along the tube, we showed that these
chiral terms change the critical behaviour in the incommensurate
phase, from
 that of a two component Luttinger liquid (in the absence of chiral interactions)
 to the usual single
component
Luttinger liquid.  However, for fields perpendicular to the axis of the tube, no chiral terms
are generated and one recovers the usual two component Luttinger liquid.
On the other hand, these terms have no effect
on
certain strongly coupled zig-zag and frustrated ladders.
For the weakly  coupled
zig-zag ladder, we  showed that the chiral term does something entirely
novel i.e., an applied magnetic field can drive a C-IC like
transition in the non-magnetic singlet sector! This results in gapless singlet and
triplet
excitations  described by a two component Luttinger liquid.
It is very remakable that   the chiral term destroys the
two component Luttinger liquid in spin tubes, but induces a
novel two component Luttinger liquid
behaviour in the weak  zig zag ladder.
These effects  should be visible in measurements of  various experimental
quantities
 ranging
from simple specific heat, NMR rates and   dynamical structure factor
measurements.

\section{Acknowledgments}
We thank F. Essler for useful discussions, A. Jezez and E. Orignac
for discussions and a careful
reading of the manuscript.

\begin{references}



\bibitem{chitra95}  D.Sen and R. Chitra, Phys. Rev. B {\bf 51}, 1992 (1995).
\bibitem{rokhsar}  D. Rokhsar, Phys. Rev. Lett {\bf 65}, 1506 (1995).

\bibitem{nersfru}  A.A. Nersesyan, A.O. Gogolin and F. H. L. Essler, Phys.
Rev. Lett. {\bf 81}, 910 (1998)

\bibitem{azaria}  P. Azaria, P. Lecheminant, A.A. Nersesyan, Phys. Rev. B
{\bf 58}, R8881 (1998)

\bibitem{andrei}  N. Andrei, M.R. Douglas, A. Jezez, Phys. Rev. B {\bf 58},
7619 (1998)

\bibitem{citro_SU(3)}  R. Citro, E. Orignac, N. Andrei, C. Itoi, J. Phys.:

\bibitem{haldane}  F.D.M. Haldane, Phys. Rev. Lett. {\bf 47}, 1840 (1981)

\bibitem{haldaneexp}  F.D.M. Haldane, Phys. Rev. Lett. {\bf 45}, 1358 (1980)

\bibitem{chaboussant}  G. Chaboussant et al. Phys. Rev. B {\bf 55}, 3046
(1997)

\bibitem{spintube}  H. Nojiri, Y. Tokunaga, M. Motokawa, Journal de Physique
49, Suppl. {\bf C8}, 1459 (1998)

\bibitem{mila_ladders}  F. Mila, Eur. Phys. J. B {\bf 6}, 201(1998)


\bibitem{white}  S.R. White and I. Affleck, Phys. Rev. B {\bf 54}, 9862
(1996)

\bibitem{mikeska_dim}  A.K. Kolezhuk and H.J. Mikeska, Int. Jour. Mod. Phys.
B {\bf 5}, 2325 (1998)

\bibitem{chitra_externalh}  R. Chitra and T. Giamarchi, Phys. Rev. B {\bf 55}%
, 5816 (1997)

\bibitem{sen}  K. Tandon and S. Lal and S. K. Pati and S. Ramasesha and D.
Sen, Phys. Rev. B {\bf 59}, 396 (1999)

\bibitem{affleck_leshouches}  I. Affleck, {\it Fields, String and Critical
phenomena} ed E. Br\.{e}zin and J. Zinn-Justin (Amsterdam. Elsevier
Science), (1988)


\bibitem{gogolin}  A.A. Gogolin, A.A. Nersesyan and A.M. Tsvelik, {\it %
Bosonization and Strongly Correlated Systems}, Cambridge University Press
(1998)

\bibitem{heinz}
H.J. Schulz, Phys. Rev. B {\bf 22},  5274 (1980).

\bibitem{zamol_extfield}  Al.B. Zamolodchikov, Int. Jour. Mod. Phys.\ {\bf A
10, }1125 (1995)

\bibitem{zamols}  S. Lukyanov and A. Zamolodchikov, Nucl. Phys. B {\bf 493,}
571 (1997)

\end{references}
\end{document}